\documentclass[conference]{IEEEtran}
\IEEEoverridecommandlockouts
\usepackage{cite}
\usepackage{amsmath,amssymb,amsfonts}
\usepackage{algorithmic}
\usepackage{graphicx}
\usepackage{textcomp}
\usepackage{xcolor}
\usepackage{indentfirst} 
\usepackage{algorithm}
\usepackage{array}
\usepackage{float}
\usepackage{bm}  
\usepackage{mathrsfs}
\usepackage{amsthm,amssymb}
\usepackage{stfloats}
\usepackage{url}
\usepackage{verbatim}
\usepackage{subfigure}
\usepackage{xfrac}  
\usepackage{titlesec}

\DeclareMathOperator{\tr}{tr}
\DeclareMathOperator{\diag}{diag}

\def\degree{${}^{\circ}$}
\def\BibTeX{{\rm B\kern-.05em{\sc i\kern-.025em b}\kern-.08em
    T\kern-.1667em\lower.7ex\hbox{E}\kern-.125emX}}
    
\begin{document}
\title{Joint Beamforming Design and 3D DoA Estimation for RIS-aided Communication System}

\author{\IEEEauthorblockN{Zhengyu Wang, Wei Yang, Tiebin Mi and Robert Caiming Qiu }
\IEEEauthorblockA{\textit{ School of Electronic Information and Communications } \\
\textit{ Huazhong University of Science and Technology, Wuhan 430074, China }\\
Email: \{wangzhengyu,yangwei\_eic,mitiebin,caiming\}@hust.edu.cn }
}

\maketitle

\begin{abstract}

In this paper, we consider a reconfigurable intelligent surface (RIS)-assisted 3D direction-of-arrival (DoA) estimation system, in which a uniform planar array (UPA) RIS is deployed to provide virtual line-of-sight (LOS) links and reflect the uplink pilot signal to sensors.
To overcome the mutually coupled problem between the beamforming design at the RIS and DoA estimation, we explore the separable sparse representation structure and propose an alternating optimization algorithm.
The grid-based DoA estimation is modeled as a joint-sparse recovery problem considering the grid bias, and the Joint-2D-OMP method is used to estimate both on-grid and off-grid parts.
The corresponding Cramér–Rao lower bound (CRLB) is derived to evaluate the estimation.
Then, the beampattern at the RIS is optimized to maximize the signal-to-noise (SNR) at sensors according to the estimated angles.
Numerical results show that the proposed alternating optimization algorithm can achieve lower estimation error compared to benchmarks of random beamforming design.

\end{abstract}


\section{Introduction}
Recently, reconfigurable intelligent surface (RIS) assisted localization has attracted growing research interest due to its potential application fields such as wireless fingerprinting localization \cite{zhang2020towards}\cite{nguyen2021wireless}, integrated sensing and communications (ISAC)\cite{wang2021joint}\cite{he2020adaptive}\cite{sankar2022beamforming}, unmanned aerial vehicle (UAV) communications\cite{chen2022efficient}, etc.
A series of research has been conducted to study the joint optimization of the active transmit beamformer at the base station (BS) and the passive reflective beamformer at the RIS in the RIS-aided localization scenarios.
For instance, the authors in \cite{song2022intelligent} adopt the Cramér–Rao lower bound as the performance metric in order to optimize the beamforming matrix based on the principle of alternating optimization.
The maximum likelihood estimation (MLE) is used to estimate the target’s DoA.
The authors in \cite{fascista2022ris} propose a reduced-complexity maximum-likelihood (ML) based estimation procedure and a codebook-based strategy for joint localization and synchronization.
In \cite{song2022joint}, the RIS was designed to maximize the minimum beampattern gain towards the sensing angles by using the techniques of alternating optimization and semi-definite relaxation (SDR).
Nevertheless, most current works ignored the beam shaping ability and high directivity of the RIS while only regarding the RIS as a signal relay or using RIS to keep the measurement matrix’s randomness.

Compress sensing (CS)\cite{donoho2006compressed} and sparse representation are often used in traditional DoA estimation problems\cite{malioutov2005sparse} with superior performance, including increased resolution and improved robustness. 
More recently, there have been many studies applying compressive sensing to RIS-aided localization systems, since the signal is usually sparse in the angular domain.
For example, the authors in \cite{chen2021reconfigurable} consider a non-uniform linear RIS array and use atomic norm-based methods (ANM) and semi-definite programming (SDP) to solve the sparse DoA estimation problem.
Similarly, a novel atomic norm-based method is proposed in \cite{chen2022ris} to remove the interference signals introduced by wireless communication. The targets' DoAs are then estimated using the Multiple Signal Classification (MUSIC) algorithms. 
Large RIS-aided near-field multipath scenarios are first proposed in  \cite{rinchi2022compressive}.
The phases of the RIS are optimized to enhance the positioning performance, and the  CS-based algorithm is used to estimate geometric parameters, including the DoA and time delay (TD).

However, there are very few works on RIS-aided DoA estimation modeling of UPA.
The expansion from 2D to 3D estimation causes a huge computational overload to traditional estimation methods such as subspace-based algorithms.
It also leads to a tremendous growth of the over-complete basis in CS problems, thus increasing the computational complexity and time.
To the best of our knowledge, this is the first work considering RIS-aided 3D DoA estimation in the far field based on UPA modeling.
The main contributions of this paper are as follows:
\begin{itemize}
\item 
First, a more general RIS-aided DoA estimation scenario is considered in 3D space. 
\item 
Second, the reduced-complexity DoA algorithm is proposed to estimate both the on-grid and grid mismatch parts. 
\item 
Third, we propose a beamforming design of RIS in the low SNR case.
\end{itemize}

\textit{Notaions:} Upper-case and lower-case boldface letters denote matrices and column vectors, respectively.
The operators $(\cdot)^T$, $(\cdot)^H$, $(\cdot)^{*}$ and $(\cdot)^{\dagger}$ denote the transpose, Hermitian transpose, conjugate, and Moore–Penrose inverse respectively.
Additionally, $\odot$, $\otimes$, and $\diamond$ represent the Hadamard matrix product, the Kronecker product, and the Khatri-Rao product, respectively.
The operator $\mathcal{R}\{\cdot \}$ denotes the real part of a complex value. $\tr(\cdot)$ is the trace of a matrix. 
$ \diag(\textbf{x}) $ returns a diagonal matrix with the elements in $\textbf{x}$ as its main diagonal entries.
$\lVert \cdot \rVert_1$ and $\lVert \cdot \rVert_2$ denote the $\ell_1$ and $\ell_2$ norm, respectively. 

\section{System Model And Problem Formulation}

\begin{figure}[t]
  \centering
  \includegraphics[width=0.85\linewidth]{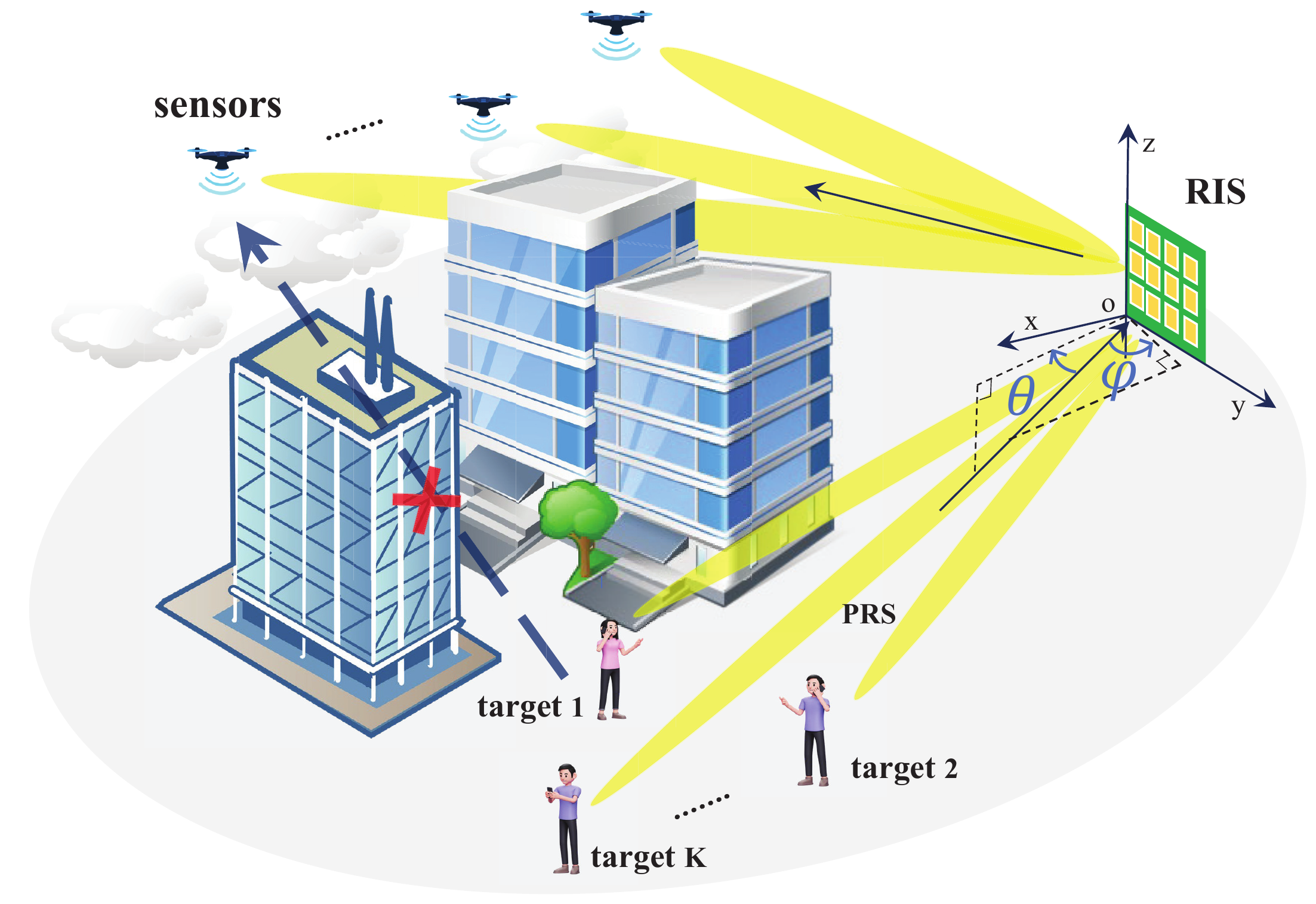}
  \caption{The system model of RIS assisted DoA estimation}\label{system model}
\end{figure}

Consider a RIS-assisted DoA estimation system, as illustrated in Fig.~\ref{system model}, where there are $R$ sensors around the BS to sense $K$ single-antenna targets.
A RIS is placed to provide virtual line-of-sight (LOS) links while the real LOS links are blocked.
We take the bottom-left element of the RIS as the origin of a 3D Cartesian coordinate system.
Assume that $K$ narrowband far-field sources impinge on the RIS from unknown directions of $(\theta_1,\phi_1),\dots,(\theta_K,\phi_K)$, where $\theta$ and $\phi$ denote the azimuth and elevation angles. 
Here elevation is the angle between the source and the xoy plane, and azimuth is the angle between the source and the yoz plane.
Considering that the RIS is a UPA with $ M \times N $ passive reflecting elements, the received signal of $(m,n)$-th element in RIS at time slot $t$ can be expressed as
\begin{equation*}
    r_{m,n}(t) = \sum_{k=1}^K e^{-j 2 \pi \frac{f}{c} \textbf{p}_{m,n}^T \textbf{u}(\theta_k,\phi_k)} x_k(t) ,
\end{equation*}
where $c$ is the speed of light and $f$ is the carrier frequency. 
$\textbf{u}(\theta_k,\phi_k)=[\sqrt{\cos\phi_k^2-\sin\theta_k^2}, \sin\theta_k, \sin\phi_k]^T$ and $\textbf{p}_{m,n} = [0,(m-1)d, (n-1)d]^T$ denote the position vector and the directional vector respectively, where $d$ is the array spacing.

Based on the separable sparse representation structure\cite{zhao2015sparse}, we divide the azimuth and elevation angles into $P$ and $Q$ grids at equal intervals and then split the manifold matrix of RIS into two individual sensing matrices
\begin{equation*}
    \textbf{A}_{\Theta} \otimes \textbf{A}_{\Phi} = [\textbf{a}_a(\theta_1),\dots,\textbf{a}_a(\theta_P)] \otimes [\textbf{a}_e(\phi_1),\dots,\textbf{a}_e(\phi_Q)] ,
\end{equation*}
where $\textbf{a}_a(\theta_k)=[e^{-j \Psi_1(\theta_k)},\dots,e^{-j \Psi_M(\theta_k)}]^T$ and 
$\textbf{a}_e(\phi_k)=[e^{-j \Psi_1(\phi_k)},\dots,e^{-j \Psi_N(\phi_k)}]^T$ denote the azimuth and elevation steering vector respectively, and $\Psi_m(\theta_k)=\frac{2 \pi}{\lambda}(m-1)d \sin \theta_k$, $\Psi_n(\phi_k)=\frac{2 \pi}{\lambda}(n-1)d \sin \phi_k$.

However, no matter how fine the grid is, the DoAs are almost surely not located exactly on the sampling grid points, which will deteriorate the performance. 
In the presence of the off-grid part, DoA $\theta_i$ (resp. $\phi_i$) can be decomposed into two parts, the nearest grid point $\Tilde{\theta_i}$ (resp. $\Tilde{\phi_i}$), and the off-grid part $\Delta^{\theta}_i$ (resp. $\Delta^{\phi}_i$), i.e. $\theta_i = \Tilde{\theta_i} + \Delta^{\theta}_i$.
Then the azimuth and elevation steering vectors can be approximated by a first-order Taylor expansion, i.e. $\textbf{a}_a(\theta_i)=\textbf{a}_a(\Tilde{\theta_i})+\textbf{b}_a(\Tilde{\theta_i})\Delta^{\theta}_i $, where $\textbf{b}_a(\Tilde{\theta_i})$ is the first-order derivative of $\textbf{a}_a(\Tilde{\theta_i})$ with respect to $\Tilde{\theta_i}$.

Suppose that the locations of $R$ sensors $(\theta_1,\phi_1),$$\dots,$ $(\theta_R,\phi_R)$ are perfectly known to the RIS.
This assumption is valid since the RIS is usually deployed at a fixed location.
Taking the grid bias into consideration, a general observation model for 3D DoA estimation at time slot $t$ can be written as
\begin{align}
    \textbf{y}_t&=\bm{\Xi}\textbf{A}_R\bm{\Omega}_t [(\textbf{A}_{\Theta}+\textbf{B}_{\Theta} \Gamma_{\alpha}) \otimes (\textbf{A}_{\Phi}+\textbf{B}_{\Phi} \Gamma_{\beta} )] \textbf{x}_t +\textbf{n}_t  \label{eq3}   \\
    &=\textbf{H}_t\textbf{x}_t+\textbf{n}_t ,  \notag
\end{align}
where $\textbf{A}_R=[\textbf{a}(\theta_1,\phi_1)^T,\dots,\textbf{a}(\theta_R,\phi_R)^T]^T \in \mathbb{C}^{R \times MN}$ represents the receiver steering matrix,
$\bm{\Xi} \in \mathbb{C}^{R \times R}$ is a diagonal matrix with the $k$-th diagonal element representing the weight parameter including all propagation effects on the signal impinging on the $k$-th sensor. 
$\textbf{n}_t \sim \mathcal{CN}(0,\sigma^2 \textbf{\textit{I}}_R)$ is the addictive white Gaussian noise vector where $\textbf{\textit{I}}_R$ is a $R \times R$ identity matrix and $\sigma$ denotes the power of the noise.
$\bm{\Omega}_t \in \mathbb{C}^{MN \times MN}$ is a diagonal phase-shift matrix of RIS, and $\textbf{H}_t \in \mathbb{C}^{R \times PQ}$ represents the channel in angular domain.
$\textbf{B}_{\Theta}=[\textbf{b}_a(\Tilde{\theta_1}),\textbf{b}_a(\Tilde{\theta_2}),\dots,\textbf{b}_a(\Tilde{\theta_P})]$ and
 $\Gamma_{\alpha}=\diag(\bm{\alpha})$, where $\bm{\alpha}=[\Delta_1^{\theta},\Delta_2^{\theta},\dots,\Delta_P^{\theta}]$. 
 The positioning reference signal(PRS) $\textbf{x}_t$ is a sparse vector with $K$ non-zero entries corresponding to the $K$ targets.  

Traditional DoA estimation methods face the problem of low precision in the case of poor SNR caused by the blocking of LOS links.  
So it is feasible to use RIS to provide an alternative reflective path and improve the SNR at sensors by properly adjusting the phase of its electromagnetic elements.
By imposing the sparsity constraint over $\textbf{x}_t$, $\bm{\alpha}$ and $\bm{\beta}$, and the received SNR constraint over $\bm{\Omega}_t$, the RIS-assisted sparsity-based DoA estimation algorithm can be formulated as
\begin{align*}
    \text{(P1)} \quad \min_{\bm{\Omega}_t, \textbf{x}_t, \bm{\alpha}, \bm{\beta}} \quad &  \lVert \textbf{y}_t - \textbf{H}_t\textbf{x}_t \rVert_2^2 + \lVert \textbf{x}_t \rVert_1 + \lambda(\lVert \bm{\alpha} \rVert_1 +\lVert \bm{\beta} \rVert_1)    \\
    \mbox{s.t.} \quad 
    & \gamma(\bm{\Omega}_t,\textbf{x}_t,\bm{\alpha},\bm{\beta}) \ge \gamma_0 \\
    & \lvert \Omega_{mn} \rvert =1, \quad mn=1,2,\dots,MN      ,
\end{align*}
where $\lambda$ is a constant controlling the tradeoff between the sparsity of the signal $\textbf{x}$ and the bias energy $(\lVert \bm{\alpha} \rVert_1 +\lVert \bm{\beta} \rVert_1)$, and $\gamma_0$ is the minimum SNR required to guarantee a valid DoA estimation.
The received SNR at sensors is expressed as $\gamma = \sfrac{\lVert \textbf{H}_t\textbf{x}_t \rVert_2^2}{\sigma^2} $.

We aim to design the optimal beamforming matrix of RIS and estimate the unknown parameters, including the DoA $\bm{\theta}, \bm{\phi}$ and corresponding off-grid parts $\bm{\alpha}$, $\bm{\beta}$, based on the received signal in \eqref{eq3}. 
However, due to the mutual coupling between the DoA estimation and the beamforming design, the problem $\text{(P1)}$ is non-convex and hard to be solved.

\section{RIS-assisted 3D DOA estimation algorithm}
In this section, we propose a RIS-assisted 3D DoA estimation method by alternatively optimizing the DoA estimation and RIS beamforming design.
Specifically, we use the Joint-2D-OMP algorithm first to estimate the angle with $\bm{\Omega}_t$ fixed, and then optimize the beamforming matrix $\bm{\Omega}_t$ by using the proposed BCDG-based manifold algorithm with $(\textbf{x}_t, \bm{\alpha}, \bm{\beta})$ fixed.
Details are as follows.

\addtolength{\topmargin}{0.01in} 

\subsection{DoA Estimation}  \label{doa estimation}

In the case of multiple snapshots, the targets send the PRS during T time slots, and RIS reflects the signal using different beamforming matrix $[\bm{\Omega}_1, \dots, \bm{\Omega}_T]$ at each time slot. Assume that $\bm{\Xi}$ and PRS remain constant during the DoA estimation phase, we drop the subscript of $\textbf{x}_t$ and the received signals are stacked into a vector as
\begin{align*}
    \begin{bmatrix}
        \textbf{y}_1  \\ \vdots \\ \textbf{y}_T
    \end{bmatrix} =
    \begin{bmatrix}
        \bm{\Xi} \textbf{A}_R \bm{\Omega}_1 \\ \vdots \\ \bm{\Xi} \textbf{A}_R \bm{\Omega}_T
    \end{bmatrix} (\textbf{A}_{\Theta}+\textbf{B}_{\Theta} \Gamma_{\alpha}) \otimes (\textbf{A}_{\Phi}+\textbf{B}_{\Phi} \Gamma_{\beta} ) \textbf{x} +
    \begin{bmatrix}
        \textbf{n}_1  \\ \vdots \\ \textbf{n}_T
    \end{bmatrix}. 
\end{align*} 

By defining $\textbf{y} = [\textbf{y}_1, \dots, \textbf{y}_T]^T$, $\textbf{n} = [\textbf{n}_1, \dots, \textbf{n}_T]^T$ and $\textbf{Z} = [\bm{\Xi} \textbf{A}_R \bm{\Omega}_1, \dots, \bm{\Xi} \textbf{A}_R \bm{\Omega}_T]^T$, the overall received signal vector can be written as
\begin{align}
    \textbf{y} = \textbf{Z} (\textbf{A}_{\Theta}+\textbf{B}_{\Theta} \Gamma_{\alpha}) \otimes (\textbf{A}_{\Phi}+\textbf{B}_{\Phi} \Gamma_{\beta} ) \textbf{x} + \textbf{n}.   \label{timestack}
\end{align}

With $\bm{\Omega}_t$ fixed, $\text{(P1)}$ can be transformed into
\begin{align*}
    \text{(P2)} \quad \min_{\textbf{x}, \bm{\alpha}, \bm{\beta}} \quad &  \lVert \textbf{y} - \textbf{Z} (\textbf{A}_{\Theta}+\textbf{B}_{\Theta} \Gamma_{\alpha}) \otimes (\textbf{A}_{\Phi}+\textbf{B}_{\Phi} \Gamma_{\beta} ) \textbf{x} \rVert_2^2 + \\
    & \lVert \textbf{x} \rVert_1 + \lambda(\lVert \bm{\alpha} \rVert_1 +\lVert \bm{\beta} \rVert_1).
\end{align*}
Note that $\text{(P2)}$ is non-convex under the constraint of both $\textbf{x}$, $\bm{\alpha}$ and $\bm{\beta}$, and a two-step process is proposed to solve the problem. 

\subsubsection{Step one}
The first term of $\text{(P2)}$ has a least squares solution given by $\textbf{Z}^{\dagger}\textbf{y}$.
By defining $\textbf{Z}^{\dagger} \textbf{y}=\text{vec}(\textbf{Y})$, we have
\begin{align}
    \text{vec}(\textbf{Y}) = (\textbf{A}_{\Theta}+\textbf{B}_{\Theta} \Gamma_{\alpha}) \otimes (\textbf{A}_{\Phi}+\textbf{B}_{\Phi} \Gamma_{\beta} ) \textbf{x}. \label{vec}
\end{align}

Solving \eqref{vec} based on the traditional Kronecker Compressed Sensing method will cause a huge computational burden because the number of atom $PQ$ is often very large. Instead, we can use joint sparse matrix reconstruction methods without converting 2D estimation into a 1D problem.

\subsubsection{Step two}
By using the property of the Kronecker product\cite{Petersen2008} $(\textbf{A}_{\Theta} \otimes \textbf{A}_{\Phi})\text{vec}(\textbf{X})=\text{vec}(\textbf{A}_{\Phi}\textbf{X}\textbf{A}_{\Theta}^T)$, and then removing the vector operator on both sides of \eqref{vec}, the equation can be expressed as
\begin{align*}
    \textbf{Y} &= (\textbf{A}_{\Phi}+\textbf{B}_{\Phi} \bm{\Gamma}_{\beta} ) \textbf{X} (\textbf{A}_{\Theta}+\textbf{B}_{\Theta} \bm{\Gamma}_{\alpha})^T \\
    &= \begin{bmatrix} \textbf{A}_{\Phi} & \textbf{B}_{\Phi} \end{bmatrix} 
    \begin{bmatrix} \textbf{X} & \textbf{X} \bm{\Gamma}_{\alpha}^T \\ \bm{\Gamma}_{\beta} \textbf{X} & \bm{\Gamma}_{\beta} \textbf{X} \bm{\Gamma}_{\alpha}^T \end{bmatrix}
    \begin{bmatrix} \textbf{A}_{\Theta}^T \\ \textbf{B}_{\Theta}^T \end{bmatrix},
\end{align*}
where $\textbf{X} \in \mathbb{C}^{Q \times P}$ and $\textbf{Y} \in \mathbb{C}^{N \times M}$. Denote that $ \textbf{P}_1=\textbf{X}, \textbf{P}_2=\textbf{X} \bm{\Gamma}_{\alpha}^T, \textbf{P}_3=\bm{\Gamma}_{\beta} \textbf{X}, \textbf{P}_4=\bm{\Gamma}_{\beta} \textbf{X} \bm{\Gamma}_{\alpha}^T $.
Since $\textbf{P}_1 \sim \textbf{P}_4$ have the same sparse structure, we propose to use Joint-2D-OMP method to estimate $\textbf{P}_1 \sim \textbf{P}_4$. The major difference of the Joint-2D-OMP algorithm is to find the maximum value of the sum of elements in the same position of $\textbf{C}^l=\textbf{A}_{\Phi}^H \textbf{R}^l \textbf{A}_{\Theta}^*$, $\textbf{D}^l=\textbf{B}_{\Phi}^H \textbf{R}^l \textbf{A}_{\Theta}^*$, $\textbf{E}^l=\textbf{A}_{\Phi}^H \textbf{R}^l \textbf{B}_{\Theta}^*$, $\textbf{F}^l=\textbf{B}_{\Phi}^H \textbf{R}^l \textbf{B}_{\Theta}^*$  in $l^{th}$ iteration, where $\textbf{R}^l$ denotes the residual error. 
Details of the Joint-2D-OMP method refer to Figure 1 in \cite{doi:10.1080/00207217.2017.1409811} 

The on-grid part can be estimated according to the value of non-zero elements in reconstructed sparse matrix $\textbf{P}_1$.
The off-grid part of the azimuth and elevation can be obtained through the non-zero elements of $\textbf{P}_1 $ divided by the elements of $\textbf{P}_2$ and $\textbf{P}_3$ at the same position, respectively.

\subsection{Optimization of Beamforming Matrix $\bm{\Omega}$}

\begin{algorithm}[tp]
  \caption{: BCGD-based Manifold Optimization Algorithm}
  \begin{algorithmic}[1]
    \REQUIRE random point $\textbf{w}_o \in \mathcal{M}$    
    \FOR{$t=1: T$}  
        \FOR{$j=1:R$}   
            \STATE using Algorithm 1 in \cite{7397861} with the cost function $-f_j(\textbf{w})=\textbf{w}^H \textbf{R}_j \textbf{w}$ and initial point $\textbf{w}_o$ as input;
            \STATE get the optimal solution $\textbf{w}_j$ on the manifold and update $\textbf{w}_o = \textbf{w}_j$;
        \ENDFOR 
    \STATE output the solution from the last iteration as $\textbf{w}_t$
    \ENDFOR
    \ENSURE  $[\textbf{w}_1, \dots, \textbf{w}_T]$
\end{algorithmic} \label{A1}
\end{algorithm}

The DoA obtained in \ref{doa estimation} can be used to guide the design of the beamforming matrix. Specifically, the receive SNR at j-th sensor is $\dfrac{\lVert \textbf{H}\textbf{x} (:,j) \rVert^2}{\sigma^2}$, where $(:,j)$ denotes the j-th row, and
\begin{align*}
    \begin{split}
        \lVert \textbf{H}\textbf{x} (:,j) \rVert^2 &= \tr[\textbf{A}_R(:,j)^H \textbf{A}_R(:,j) \bm{\Omega} (\textbf{A}_u \textbf{x})(\textbf{A}_u \textbf{x})^H \bm{\Omega}^H] \\
        &= \tr(\bm{\Omega}^H \textbf{B}_j \bm{\Omega} \textbf{C}) ,
    \end{split}
\end{align*}
where we use the property $\tr(\bm{\Omega}^H \textbf{B}_j \bm{\Omega} \textbf{C})=\textbf{w}^H (\textbf{B}_j \odot \textbf{C}^T) \textbf{w}$,
and the notation
$\textbf{A}_u=(\textbf{A}_{\Theta}+\textbf{B}_{\Theta} \Gamma_{\alpha}) \otimes (\textbf{A}_{\Phi}+\textbf{B}_{\Phi} \Gamma_{\beta} )$ ,$\textbf{B}_j=\textbf{A}_R(:,j)^H \textbf{A}_R(:,j)$, $\textbf{C}=(\textbf{A}_u \textbf{x})(\textbf{A}_u \textbf{x})^H$ for simplicity.

It is more reasonable to maximize the SNR at each sensor than maximize the overall SNR because the maximum $\ell_2$ norm of $ \textbf{H}\textbf{x}$ does not mean that each element in $ \textbf{H}\textbf{x}$ is maximized equally.
Different from the method in existing literature, we model the beamforming problem as an optimization problem of R copies of $\textbf{w}$ with $(\textbf{x}, \bm{\alpha}, \bm{\beta})$ fixed.
\begin{align*}
    \text{(P3)} \quad \max_{\textbf{w}^{(j)}} \quad & \sum_{j=1}^R(\textbf{w}^{(j)})^H \textbf{R}_j \textbf{w}^{(j)} \quad j=1, \dots, R \\
    \mbox{s.t.} \quad 
    & \lvert w_{i}^{(j)} \rvert =1, \quad i=1,2,\dots,MN  \\
    & \textbf{w}^{(1)}=\textbf{w}^{(2)}=\dots=\textbf{w}^{(R)} ,
\end{align*}
where $\textbf{R}_j=\textbf{B}_j \odot \textbf{C}^T$ and $\textbf{w}^{(j)}\in \mathbb{C}^{MN \times 1}$ is a vector formed by the diagonal elements of $\bm{\Omega}$.
Note that $\text{(P3)}$ can be improved to weighted summation when there is a large difference between the performance of the sensors.

To solve $\text{(P3)}$, we proposed a block coordinate gradient descent(BCGD) based manifold optimization algorithm as shown in Algorithm \ref{A1}.
For each subproblem $f_j(\textbf{w})=\textbf{w}^H \textbf{R}_j \textbf{w}$, 
the unit-modulus constrained problem in Euclidean space can be transformed into an unconstrained problem on complex circle manifolds $\mathcal{M}$ \cite{HU2022108322}\cite{8706630} and then the Riemann gradient steepest descent algorithm is used to solve the optimization problem.

\addtolength{\topmargin}{0.01in} 

We aim to maximize $R$ subproblems alternatively using manifold optimization and update the solution of the previous subproblem to the next initial point on the manifold. 
Besides, we can set the maximum iteration time relatively small, which significantly reduces the complexity and time of manifold optimization. 

The overall algorithm for joint beamforming and the DoA estimation is shown in Algorithm \ref{A2}.

\begin{algorithm}[tp]
  \caption{: Overall Algorithm}
  \begin{algorithmic}[1]
    \REQUIRE random beamforming matrix of RIS
    \REPEAT
    \STATE the targets transmit a starting signal to the RIS and sensors for synchronization,
    \STATE the targets start to transmit PRS, and RIS changes its configuration sequentially denoted by $[\bm{\Omega}_1, \dots, \bm{\Omega}_T]$ obtained according to Algorithm \ref{A1},
    \STATE all received signals are stacked into a new vector $\textbf{y}$ and then sent to the BS for DoA estimation using the method proposed in \ref{doa estimation}
    \STATE updates the DoAs for the guidance of RIS configuration in the next iteration.
    \UNTIL{the value of DoAs converged}
    \ENSURE  the targets' DoAs
\end{algorithmic} \label{A2}
\end{algorithm}

\subsection{Complexity Analysis}
The complexity of our proposed DoA estimation method is $\mathcal{O}((MN)^3)+\mathcal{O}((KMNPQ))$ in each iteration, where $K$ is the number of targets, $P, Q$ are the number of grids which are usually on the same order of magnitude as $N, M$.
So the complex order of the proposed method is about $\mathcal{O}((MN)^3)+\mathcal{O}(K(MN)^2)$ per iteration, while the complexity of the conventional atomic norm-based passive DoA estimation method is $\mathcal{O}((MN)^4)+\mathcal{O}((MN)^3)$.


\section{SIMULATION RESULTS}
In the following simulations, we consider a distance-dependent path loss $\rho=10^{-3}(\frac{d_t}{d_0})^{- \alpha_0}$, where $d_t$ is the distance of the transmission link and the reference distance $d_0$ is set as $1m$. 
\begin{figure}[b]
\centering
	\subfigure[iteration=1]{
    \includegraphics[width=0.46\linewidth]{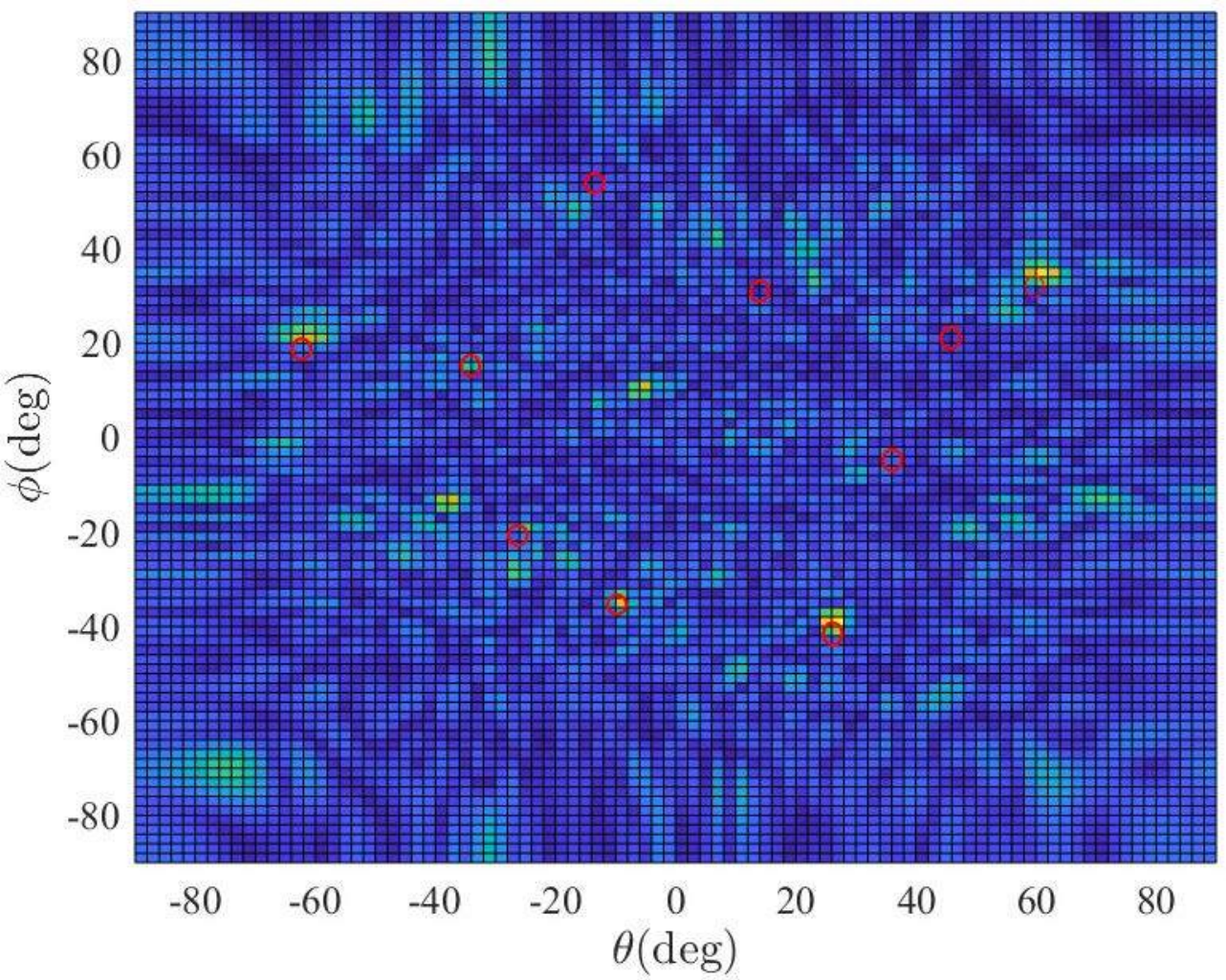}
    }\subfigure[iteration=10]{
    \includegraphics[width=0.45\linewidth]{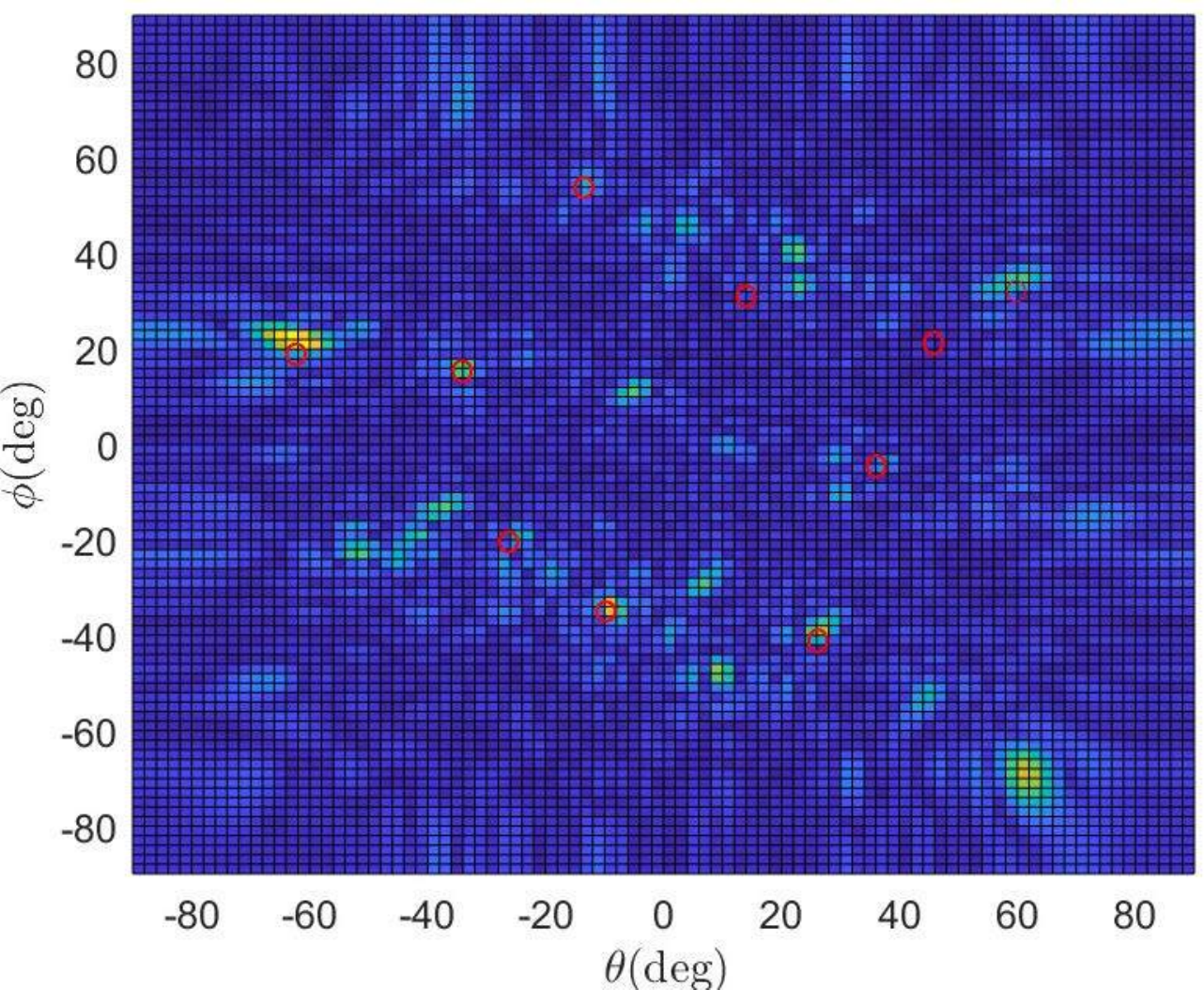}
    }
 \caption{Output field of RIS at iterations 1(a) and 10(b). The optimization is solved by the Manopt toolbox, and the maximum iteration time is set as 10. }
 \label{outputRIS}
\end{figure}
The path exponent $\alpha_0$ is set as 2.2. 
Assume that all paths share the same path loss weight, $\Xi$ is equal to $\rho \textit{I}$.
The number of sensors, targets, and reflecting elements is set to be $R=10$, $K=6$, and $M=N=20$ respectively.
The array spacing of RIS is set to half of the
wavelength. 
A uniform sampling grid is applied over the angular space $[-90^{\circ},90^{\circ]}$ in both the elevation and azimuth directions with grid intervals of $3^{\circ}$, so the grid number $P=Q=61$. 
Assume all PRSs have equal power and are set to 1. 
The estimation performance is measured by the RMSE defined as $\text{RMSE}=\frac{1}{2} (\text{RMSE}_{\theta} + \text{RMSE}_{\phi} )$, 
and the RMSE of the azimuth and elevation angle are
\begin{align*}
    \text{RMSE}_{\theta} &= \sqrt{\frac{1}{N_m K} \sum_{i=1}^{N_m}\sum_{k=1}^K(\hat{\theta}_{i,k}-\theta_k )^2} \\
    \text{RMSE}_{\phi} &= \sqrt{\frac{1}{N_m K} \sum_{i=1}^{N_m}\sum_{k=1}^K(\hat{\phi}_{i,k}-\phi_k )^2} ,
\end{align*}
where we set the Monte Carlo simulation times $N_m = 100$.
\begin{figure}[t]
  \centering
  \includegraphics[width=0.7\linewidth]{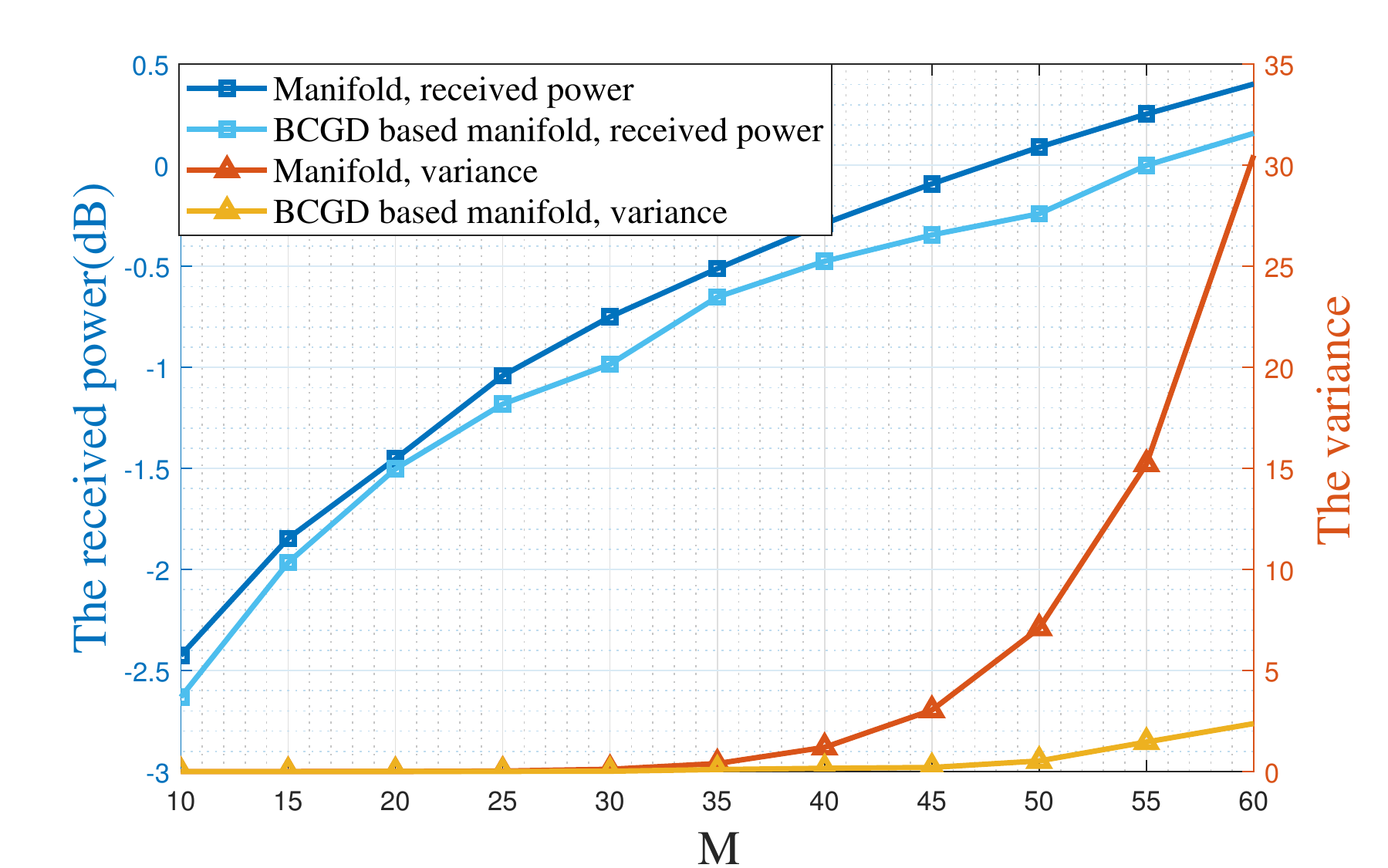}
  \caption{The receiving power (left axis) and variance (right axis) versus $M$ using our proposed BCGD-based manifold optimization and ordinary manifold optimization, and we assume that $N=M$.
  }\label{manifold}
\end{figure}
\begin{figure}[t]
  \centering
  \includegraphics[width=0.7\linewidth]{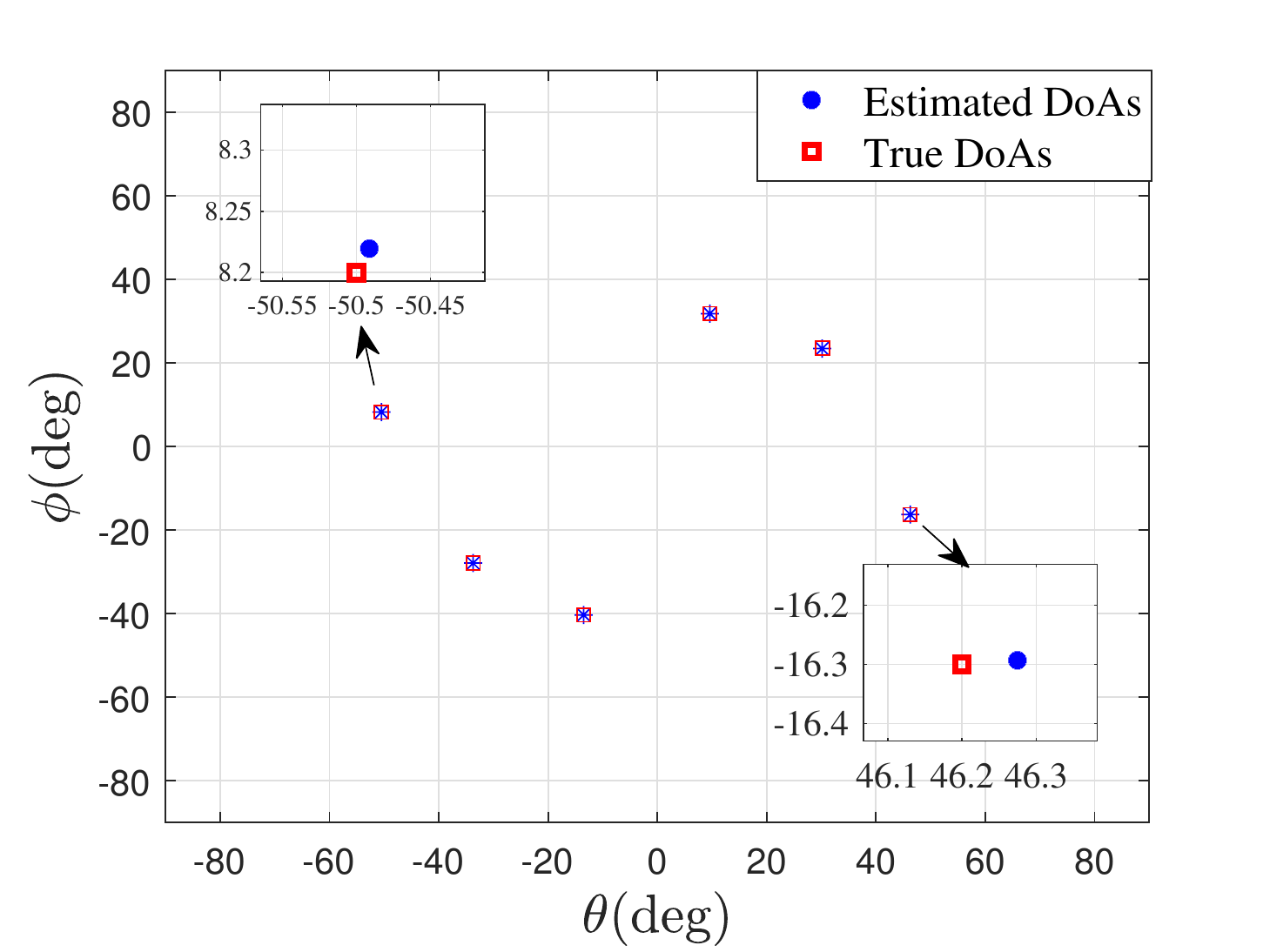}
  \caption{The ground truth DoA versus the estimated DoA.} \label{doa result}
\end{figure}

First, we design the beamforming matrix of the RIS according to Algorithm \ref{A1} with the optimization objective of maximizing the SNR of each sensor. 
We consider 6 targets from directions $(-50.5^{\circ},8.2^{\circ})$,$(-33.7^{\circ},-27.9^{\circ})$,$(-13.5^{\circ},-40.3^{\circ})$, $(9.6^{\circ},31.8^{\circ})$ ,$(30.2^{\circ},23.5^{\circ})$, $(46.2^{\circ},-16.3^{\circ})$.
The output field of the RIS at different iteration times is shown in Fig.~\ref{outputRIS}, and the coordinates of sensors are marked with red circles.
We can observe that the power is concentrated toward sensors as the iteration times increase.

Second, we compared the receiving power and its variance under different numbers of RIS elements $M$, 
as shown in Fig.~\ref{manifold}.
The results show that our proposed BCGD-based manifold optimization can effectively reduce the variance between the received powers while ensuring the optimal solution and low complexity.

In Fig.~\ref{doa result}, we show the performance of the off-grid DoA estimation with the optimized RIS under SNR=10dB. 
It can be found that the algorithm can jointly recover the on-grid angles and the off-grid parts. 
Specifically, take the target from (46.2\degree,-16.3\degree) as an example, the estimation error for the azimuth and elevation are $0.0747$ and $0.0068$, respectively, as illustrated in the bottom right corner of Fig.~\ref{doa result}.

Then, we compare the RMSE of our proposed method with three benchmark schemes:
\begin{itemize}
\item \textit{the proposed method while under the random RIS setting:}
The beamforming matrix of the RIS is randomly set at each time slot. 

\item \textit{the proposed method while the RIS is optimized to minimize CRLB:}
the formulation of CRLB is derived in Appendix~\ref{derivation of CRB}. 
The CRLB minimization can be transformed into a quadratic optimization problem.
For details, refer to Appendix~\ref{optimRIS_CRB}.

\item \textit{ANM method under optimized RIS settings:}
ANM is a continuous angle domain search method.
The sparse reconstruction problem can be transformed into an SDP problem \cite{chen2022ris} and then solved by CVX toolbox \cite{grant2014cvx} in MATLAB.
\end{itemize}



\begin{figure}[t]
  \centering
  \includegraphics[width=0.75\linewidth]{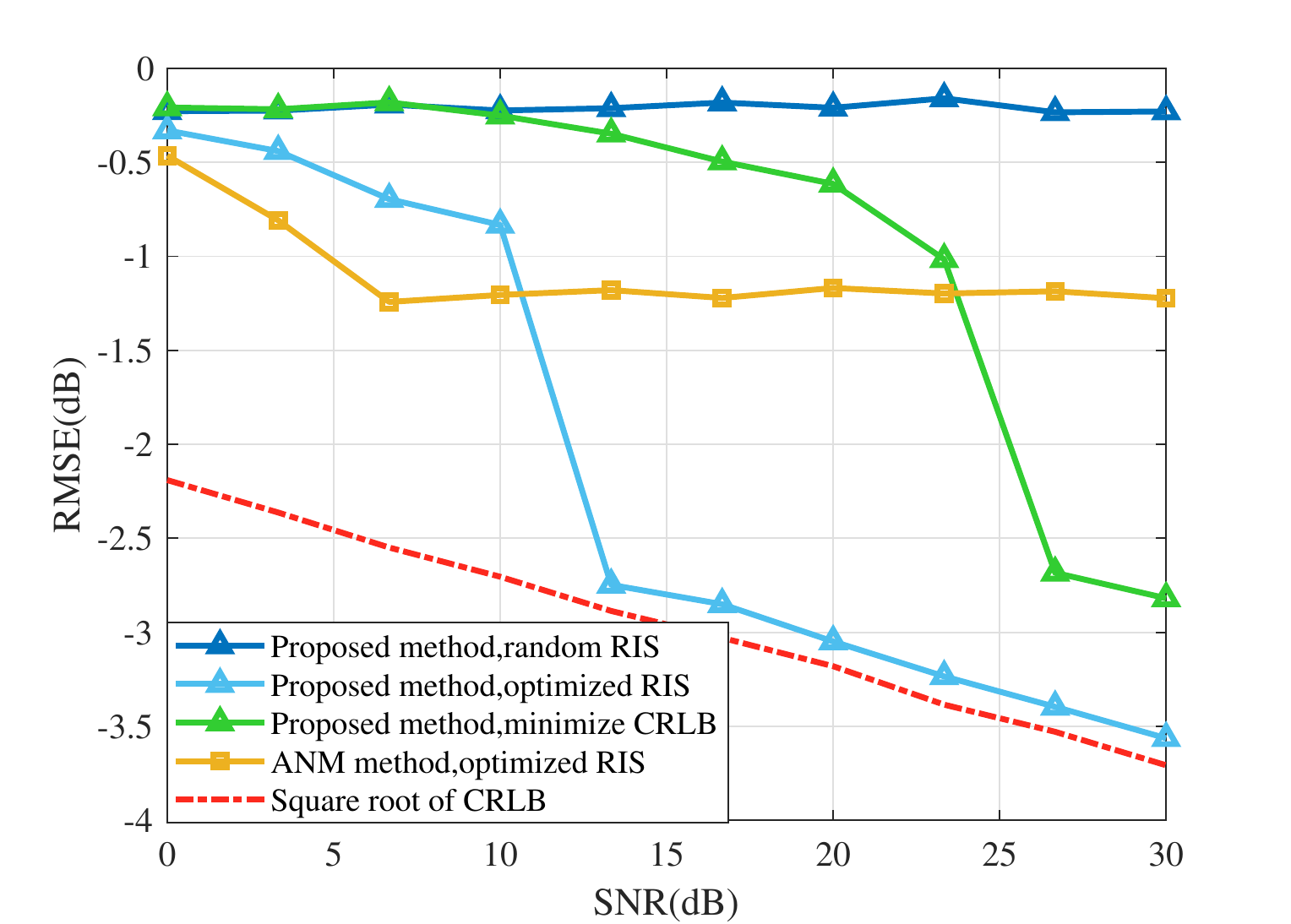}
  \caption{RMSE versus SNR using different methods. }\label{final}
\end{figure}

Note that the optimized RIS is used in the calculation of CRLB.
We can see from Fig.~\ref{final} that compared with random RIS, lower RMSE can be achieved with the optimized RIS.
Moreover, the RIS optimized with the goal of maximizing the SNR performs better in the low SNR case, as compared to the RIS optimized with the goal of minimizing the CRLB.

\begin{figure}[t]
  \centering
  \includegraphics[width=0.7\linewidth]{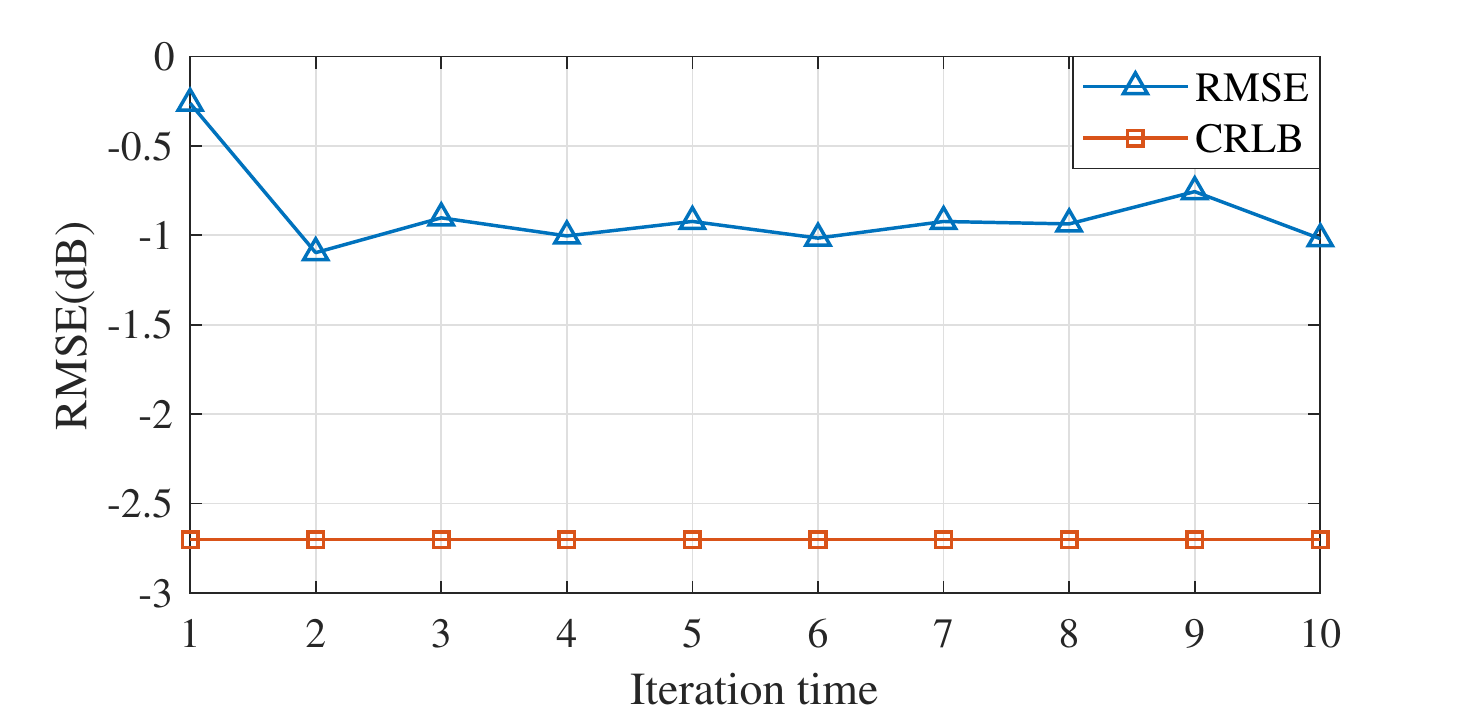}
  \caption{ RMSE versus iteration times.}\label{iter}
\end{figure}

Finally, the convergence of the proposed algorithm is shown in Fig.~\ref{iter}.
The initial SNR is set to 10dB, and the alternating optimization procedure refers to Algorithm 2.
We can find that the RMSE of DoA estimation gradually decreases and tends to converge with a fluctuation of fewer than 0.1 degrees after the second iteration.

\section{CONCLUSION}
To solve the coupled RIS-assisted DoA estimation problem, we proposed an alternating optimization algorithm by jointly optimizing the DoA estimation and RIS beamforming design.
Using the separable sparse representation structure, the azimuth and elevation angles can be jointly estimated by the reduced-complex Joint-2D-OMP algorithm.
Subsequently, a BCGD-based manifold optimization algorithm was proposed for RIS beamforming design.
Simulation results validated the effectiveness of the proposed algorithm.

\begin{appendices}
\section{}  \label{derivation of CRB}
Appendix \ref{derivation of CRB} gives the derivation of CRLB.
The non-sparse form of eq. \eqref{timestack} is equivalent to
\begin{align*}
    \textbf{y} = \textbf{Z} (\textbf{A}_{\theta} \diamond \textbf{A}_{\phi}) \textbf{s} + \textbf{n} ,  
\end{align*}
where $\textbf{A}_{\theta} = [ \textbf{a}_a(\theta_1), \dots, \textbf{a}_a(\theta_K) ]$ and $\textbf{A}_{\phi} =  [ \textbf{a}_e(\phi_1), \dots, \textbf{a}_e(\phi_K) ]$ are matrices of steering vectors for true azimuth and elevation angles, respectively. $\textbf{s}\in \mathbb{C}^{K \times 1}$ is the signal sent by $K$ targets.

The probability density function of the received signal can be expressed as 
\begin{align*}
    f(\textbf{y};\bm{\xi}) = \frac{1}{\pi^R \det(\bm{\Sigma})} e^{-(\textbf{y}-\bm{\mu})^H \bm{\Sigma}^{-1} (\textbf{y}-\bm{\mu})} ,
\end{align*}
where the mean and covariance matrix are
\begin{align*}
    \bm{\mu} &= \textbf{Z} (\textbf{A}_{\theta} \diamond \textbf{A}_{\phi}) \textbf{s},  \\
    \bm{\Sigma} &= \sigma^2 \textit{\textbf{I}}_R .
\end{align*} 

To obtain the CRLB, we first derive the Fisher information matrix (FIM) $\textbf{F}$ by calculating the partial derivative of the probability density function with respect to parameters. 
Collect all the unknown parameters into a vector $\bm{\xi}=[\bm{\theta}^T , \bm{\phi}^T ]^T$, then
\begin{flalign*}
    \frac{\partial \ln f(\textbf{y};\bm{\xi})}{\partial \bm{\theta}} = (\textbf{y}-\bm{\mu})^H \bm{\Sigma}^{-1} \frac{\partial \textbf{Z}(\textbf{A}_{\theta} \diamond \textbf{A}_{\phi}) \textbf{s}}{\partial \bm{\theta}}+
    [\bm{\Sigma}^{-1} \\
    (\textbf{y}-\bm{\mu})]^T \frac{\partial [\textbf{Z} (\textbf{A}_{\theta} \diamond \textbf{A}_{\phi}) \textbf{s}]^{*}}{\partial \bm{\theta}}  \\
    = 2 \mathcal{R} \Big\{ (\textbf{y}-\bm{\mu})^H \bm{\Sigma}^{-1} \frac{\partial \textbf{Z} (\textbf{A}_{\theta} \diamond \textbf{A}_{\phi}) \textbf{s}}{\partial \bm{\theta}} \Big\}  \\
    = \frac{2}{\sigma^2} \mathcal{R} \Big\{ (\textbf{y}-\bm{\mu})^H \textbf{Z}(\dot{\textbf{A}_{\theta}} \diamond \textbf{A}_{\phi}) \diag(\textbf{s}) \Big\} ,
\end{flalign*}
where $ \dot{\textbf{A}_{\Theta}} = [\dot{\textbf{a}(\theta_1)},\dot{\textbf{a}_a(\theta_2)},\dots,\dot{\textbf{a}_a(\theta_K)}]$, 
and $\dot{\textbf{a}(\theta_i)}$ denotes the derivative of steering vector $\textbf{a}(\theta_i)$ with respect to $\theta_i$. 

Similarly,
\begin{align*}
    \frac{\partial \ln f(\textbf{y};\bm{\xi})}{\partial \bm{\phi}} &= \frac{2}{\sigma^2} \mathcal{R} \Big\{(\textbf{y}-\bm{\mu})^H \textbf{Z} (\textbf{A}_{\theta} \diamond \dot{\textbf{A}_{\phi}}) \diag(\textbf{s}) \Big\}.
\end{align*}

Then the FIM matrix $\textbf{F}$ can be partitioned as
 \begin{align*}
     \textbf{F} = \begin{bmatrix}
        \textbf{F}_{\bm{\theta} \bm{\theta}}  &  \textbf{F}_{\bm{\theta} \bm{\phi}} \\
        \textbf{F}_{\bm{\phi} \bm{\theta}}  &  \textbf{F}_{\bm{\phi} \bm{\phi}}
     \end{bmatrix} = \frac{2 \rho^2}{\sigma^2} \mathcal{R} \begin{bmatrix}
     \bm{\Lambda}^H \bm{\Lambda} &  \bm{\Lambda}^H \bm{\Upsilon} \\
     \bm{\Upsilon}^H \bm{\Lambda} &  \bm{\Upsilon}^H \bm{\Upsilon} 
     \end{bmatrix} ,
 \end{align*}
where we define
\begin{align*}
    \bm{\Lambda} = \textbf{Z} (\dot{\textbf{A}_{\theta}} \diamond \textbf{A}_{\phi}) \text{diag}(\textbf{s}) \\
    \bm{\Upsilon} = \textbf{Z} (\textbf{A}_{\theta} \diamond \dot{\textbf{A}_{\phi}}) \text{diag}(\textbf{s}) .
\end{align*}

Therefore the CRLB for DoAs is given as $\text{CRLB}=\tr[\textbf{F}^{-1}]$.

\section{} \label{optimRIS_CRB}
Appendix \ref{optimRIS_CRB} gives the derivation of the RIS beamforming matrix for minimizing CRLB in the second benchmark scheme.

By defining $\textbf{D}_1 = \dot{\textbf{A}_{\theta}} \diamond \textbf{A}_{\phi}$, $\textbf{D}_2 = \textbf{A}_{\theta} \diamond \dot{\textbf{A}_{\phi}}$ and $\textbf{P}_{i,j} = s_i^{*}s_j$, we have
\begin{align*}
    \textbf{F} &= \frac{2 \rho^2}{\sigma^2} \mathcal{R} \Big\{ \begin{bmatrix}
     (\textbf{D}_1^H \textbf{Z}^H \textbf{Z} \textbf{D}_1)\odot \textbf{P} &  (\textbf{D}_1^H \textbf{Z}^H \textbf{Z} \textbf{D}_2)\odot \textbf{P} \\
     (\textbf{D}_2^H \textbf{Z}^H \textbf{Z} \textbf{D}_1)\odot \textbf{P} &  (\textbf{D}_2^H \textbf{Z}^H \textbf{Z} \textbf{D}_2)\odot \textbf{P} 
     \end{bmatrix} \Big\} \\
     &=\frac{2 \rho^2}{\sigma^2} \mathcal{R} \Big\{ \begin{bmatrix} \textbf{D}_1^H \\ \textbf{D}_2^H \end{bmatrix} \textbf{Z}^H\textbf{Z}  \begin{bmatrix} \textbf{D}_1 & \textbf{D}_2\end{bmatrix}  \Big\} ,
\end{align*} 
where we can assume that $\textbf{P}$ is an all-ones matrix. 
Define $\textbf{D}=\begin{bmatrix} \textbf{D}_1 & \textbf{D}_2\end{bmatrix}$, then
\begin{align*}
     \textbf{F} &= \frac{2 \rho^2}{\sigma^2} \mathcal{R} \Big\{ 
     \begin{bmatrix} \textbf{D}^H\bm{\Omega}_1^H \dots \textbf{D}^H\bm{\Omega}_T^H \end{bmatrix} (\bm{\Xi} \textbf{A}_R)^H(\bm{\Xi} \textbf{A}_R) 
     \begin{bmatrix} \textbf{D}\bm{\Omega}_1 \\ \vdots \\ \textbf{D}\bm{\Omega}_T \end{bmatrix} \Big\} \\
     &=\frac{2 \rho^2}{\sigma^2} \mathcal{R} \Big\{ \begin{bmatrix} \textbf{D}^H \dots \textbf{D}^H\ \end{bmatrix} \bm{\Omega}_{d}^H \textbf{Q} \bm{\Omega}_{d} \begin{bmatrix} \textbf{D}^H \dots \textbf{D}^H \end{bmatrix}^H \Big\} \\
     &=\frac{2 \rho^2}{\sigma^2} \mathcal{R} \Big\{ \textbf{B}^H \bm{\Omega}_d^H \textbf{Q} \bm{\Omega}_d \textbf{B} \Big\}  \\
     &=\frac{4 \rho^2}{\sigma^2} \textbf{B}^H \bm{\Omega}_d^H \textbf{Q} \bm{\Omega}_d \textbf{B} ,
\end{align*}
where $\textbf{B} = \begin{bmatrix} \textbf{D}^H \dots \textbf{D}^H \end{bmatrix}^H$,  $\bm{\Omega}_d = \diag(\bm{\Omega}_1, \dots, \bm{\Omega}_T)$, and $\textbf{Q}$ returns a blocked diagonal matrix with $(\bm{\Xi} \textbf{A}_R)^H(\bm{\Xi} \textbf{A}_R) $ as elements.

Then the CRLB is expressed as
\begin{align*}
    \text{CRLB} &= \tr(\textbf{F}^{-1}) = \frac{4 \rho^2}{\sigma^2} \tr[\bm{\Omega}_d^H (\textbf{B}\textbf{B}^H)^{-1} \bm{\Omega}_d \textbf{Q}^{-1}] \\
    &= \frac{4 \rho^2}{\sigma^2} \bm{\omega}^H [(\textbf{B}\textbf{B}^H)^{-1} \odot (\textbf{Q}^{-1})^T] \bm{\omega} ,
\end{align*}
where $\bm{\omega}$ is a vector formed by the diagonal elements of $\bm{\Omega}_d$. 
The CLRB minimization problem is equivalent to a quadratic optimization problem, which can be solved by SDP or manifold optimization.

\end{appendices}

\vspace{12pt}

\bibliographystyle{IEEEtran}

\end{document}